\documentstyle[12pt]{article}
\begin{document}

\begin{center}
{\Large\bf Hyperscaling in the Domany-Kinzel Cellular Automaton}
\\[0.5\baselineskip]
\vspace{2em}

{\large Ronald Dickman$^*$}\\[0.5\baselineskip]
Department of Physics and Astronomy\\
Lehman College, CUNY\\
Bronx, NY 10468-1589, USA\\[0.5\baselineskip]
and\\[0.5\baselineskip]
{\large Alex Yu. Tretyakov$^{\dag}$}\\[0.5\baselineskip]
Graduate School of Information Sciences\\
Tohoku University\\
Aoba-ku, Sendai 980, Japan\\[0.5\baselineskip]
\end{center}
\vspace{1em}

\begin{abstract}
An apparent violation of hyperscaling at the endpoint of the critical
line in the Domany-Kinzel stochastic cellular automaton finds an elementary
resolution upon noting that the order parameter is discontinuous at this
point.  We derive a hyperscaling relation for such transitions and
discuss applications to related examples.
\end{abstract}
\vspace{1em}

$^*$ e-mail address: dickman@lcvax.lehman.cuny.edu\\
$^{\dag}$ e-mail address: alex@fractal.mech.tohoku.ac.jp\\

\newpage

Domany and Kinzel's stochastic cellular automaton (DKCA)
continues to attract interest
as a particle system affording
a test of universality in
nonequilibrium critical phenomena.\cite{domany,kinz}
In the phase diagram of the DKCA, a line of critical points
separates the active phase from the absorbing vacuum phase.
There is good numerical evidence \cite{kinz,kohring,bagnoli}
that the critical behavior
along this line is that of directed percolation (DP), except at
one terminal point (called {\em compact directed percolation} or CDP),
where the asymptotic behavior is known exactly.
Essam showed that the critical exponents
governing the
moments of the cluster-size distribution in CDP
violate the usual hyperscaling relation, and derived
the appropriate generalization.\cite{essam}
In this note we point out that the hyperscaling
relation amongst the exponents describing spreading also
requires modification.   This
leads us to formulate a hyperscaling relation for
{\em discontinuous} transitions into an absorbing state.

In the one-dimensional DKCA, the state $\sigma_{\rm i}(t+1)$ (=0 or 1), of
site i at time $t+1$ depends upon $T_{\rm i} \equiv
\sigma_{\rm i}(t) + \sigma_{\rm i-1}(t)$.  If we let
$\Pr [\sigma_{\rm i}(t+1) = 1$]
= $h(T_{\rm i})$,
then $h(0) = 0$, $h(1) = p_1$, and $h(2) = p_2$.
(All sites are updated simultaneously in this discrete-time process.)
For each
$p_2 < 1$ there is a critical value $p_{1,c}$, such that $\delta_{\phi}$
(the vacuum --- all sites in state 0), is
the unique steady state for $p_1 < p_{1,c}$, whilst for $p_1 > p_{1,c}$ there
is
also an ``active" stationary state with a nonzero density of sites in state 1.
It is believed that the critical behavior along the line $p_{1,c} (p_2)$ is
the same as that of DP, for $p_2 < 1$.

The case $p_2 = 1$, (CDP) is special.  On this line, $\delta_1$, the all-1
state,
is absorbing, as well as $\delta_\phi$. The evolution
is determined by the coalescence of random walks marking the
boundaries between strings of 0's and 1's; for
$p_1 = 1/2$ the walks are {\em unbiased}.
In fact, ($p_1 = 1/2,p_2=1)$ is a
critical endpoint:
for $p_1 < 1/2$, $\delta_\phi$ is the globally attractive stationary state, and
conversely for $\delta_1$, when $p_1 > 1/2$.  (By ``globally attractive"
we mean that
for $p_1 < 1/2$, initial configurations with an infinite number of 0's
evolve to $\delta_\phi$
with probability 1, and that there is a nonzero probability to reach
$\delta_{\phi}$ if there is at least one
site in state 0 initially.)  Let $p_1 = 1/2 + \Delta$.
It is easy to see that CDP is invariant under
the transformation which takes $\Delta$ to
$-\Delta$ and interchanges 0's and 1's.

Consider the evolution at (1/2,1), starting
with the origin in state 1, all other sites zero.
The number $n(t)$ of 1's at time $t$ is a simple
random walk on {\bf Z} starting at $n(0) = 1$, with an absorbing boundary
at the origin.  From well-known
properties of random walks, we have that the survival probability
$P(t) = \Pr[n(t') > 0, \; t' \leq t] \sim t^{-1/2}$,
and that the mean-square population of 1's in {\em surviving} trials is
$n_{surv}(t) = $ E[$n(t)^2 | n(t') > 0, \; t' \leq t] \sim t$.  It follows
that the mean number of 1's (averaged over all trials, including those
which die out)
is asymptotically {\em O}(1).
Spreading of a critical process from a localized
source is customarily described by a
set of exponents defined {\em via} the relations \cite{torre}

\begin{equation}
P(t) \propto t^{-\delta},
\end{equation}
\begin{equation}
n(t) \propto t^{\eta},
\end{equation}
and
\begin{equation}
R^2(t) \propto t^z ,
\end{equation}

\noindent where $n(t)$ denotes the mean population size
and $R^2(t)$ the mean-square
spread of particles (1's, in our notation) about the origin.
Along the critical line in the DKCA, the exponents $\delta$, $\eta$, and $z$
take universal values ({\em i.e.}, those of 1+1-dimensional DP), {\em except}
at
($1/2,1$) where evidently $\delta = 1/2$, $\eta = 0$, and $z=1$.

The spreading exponents are connected by a hyperscaling relation \cite{torre}
\begin{equation}
4 \delta + 2 \eta = dz,
\label{hyper}
\end{equation}
where $d$ is the number of spatial dimensions.
While the exponents for DP are consistent with this relation,
those for the point (1/2,1) (with $d=1$) aren't.  Does this signal
a violation of hyperscaling in CDP?
Here it is useful to note that Eq (\ref{hyper})
is a special case of a more general relation\cite{mendes}
\begin{equation}
\label{ghyper}
(1 + \frac{\beta}{\beta'}) \delta +  \eta = \frac{dz}{2},
\end{equation}
where $\beta $ is the usual order-parameter exponent,
defined, for the DKCA, through
$\rho_1 \propto
(p_1 - p_{1,c})^{\beta}$, for
$p_1 > p_{1,c}$;
$\rho_1$ is the stationary density of 1's.
$\beta' $ governs the ultimate survival probability (starting from a localized
source): $P_{\infty} \equiv \lim_{t \rightarrow \infty} P(t) \propto
(p_1 - p_{1,c})^{\beta'}$.
Eq (\ref{ghyper}) reduces to Eq(\ref{hyper}) when the order parameter
and $P_{\infty}$ are governed by  the same exponent, as they are
in the contact process and other DP-like models.  Models with
multiple absorbing configurations \cite{mendes},
and branching annihilating random walks with even parity \cite{iwan94},
can have $\beta \neq \beta'$, in which case
 the exponents violate Eq (\ref{hyper}), but satisfy Eq (\ref{ghyper}).

It is known that $\beta' = 1$ in CDP. \cite{domany,essam,durrett,bnote}
The order-parameter exponent, $\beta$,
by contrast, is {\em zero}.  This is because (1/2,1)
marks a {\em discontinuous} transition, as is readily seen by
recalling the symmetry property noted above.  Since $\rho_1 = 0$ for
$p_1 < 1/2$,
(the globally attractive state is $\delta_{\phi}$), it follows
that $\rho_1 = 1$ for
$p_1 > 1/2$.  Strictly speaking, $\beta$ is not defined here.
But since $lim_{\beta \searrow 0} \; \Delta^{\beta} = 1$
for any $\Delta > 0$, it is natural to associate the value $\beta = 0$ with
the discontinuous transition.  Indeed, the values $\delta = 1/2$, $z=d=1$,
and $\eta = \beta = 0$ yield an identity when inserted in Eq (\ref{ghyper}).

In fact, we can eliminate any reference to the ill-defined exponent $\beta$
by adapting Grassberger and de la Torre's scaling argument
to discontinuous transitions.  Consider a model
with a transition from an absorbing to
an active state at $\Delta = 0$, with exponents $\delta$, $\eta$, $z$,
and $\beta'$
defined as above.
Suppose, however, that the order parameter $\rho$ is discontinuous, being
zero for $\Delta < 0$, and
\begin{equation}
\rho = \rho_0 + f(\Delta),
\end{equation}
for $\Delta > 0$, where $\rho_0 > 0$, and $f$ is continuous and vanishes
at $\Delta = 0$.  The scaling hypothesis for spreading from a source
postulates the existence of two scaling functions, defined {\em via}
\cite{torre}

\begin{equation}
\rho(x,t) \sim t^{\eta - dz/2} G(x^{2}/t^{z}, \Delta t^{1/\nu_{||}}),
\end{equation}
and
\begin{equation}
\label{gensp}
P(t) \sim t^{-\delta} \Phi(\Delta t^{1/\nu_{||}}).
\end{equation}
(Here $\rho(x,t)$ is the local order-parameter density.
$\nu_{||}$ governs the divergence of the correlation time $\tau$ at the
transition: $\tau \sim \Delta^{-\nu_{||}}$.)  Existence of the limit
$P_{\infty}$
implies that $\Phi (x) \sim x^{\beta'} $ as $x \rightarrow \infty$,
with $\beta' = \delta \nu_{||}$.  In a surviving trial, the local
density must approach the stationary density $\rho$ as $t \rightarrow \infty$,
so $\rho(x,t) \sim \Delta^{\beta'} \rho_0 $,
for $t \rightarrow \infty$ with fixed $x$, and $\Delta $ small but positive.
It follows that $G(0,y) \sim y^{\beta'} $ for large $y$.
On the other hand, we must have $G(0,y) \sim y^{-\nu_{||}(\eta - dz/2)} $
for $\lim_{t \rightarrow \infty} \rho(x,t) $ to exist.  Comparing
these asymptotic behaviors, we find a
hyperscaling relation for transitions at which the order parameter
is discontinuous

\begin{equation}
\label{dhyper}
\delta + \eta = \frac {dz}{2}.
\end{equation}

\noindent The interpretation is immediate:
simply note that
$\delta + \eta$ is the exponent governing the mean population
in surviving trials, and that the radius $R_t$ of such a cluster
grows $ \sim t^{z/2}$.
(Since the density is positive, clusters are compact, not
fractal as in DP.)  Eq (\ref{dhyper}) is nothing more than
the scaling law for the volume of a
$d$-dimensional sphere of radius $R_t$.
It should apply whenever power-law growth produces {\em
compact colonies}.  (By ``colony" we mean the set of
particles, or 1's --- in general, active sites --- at time $t$,
descended from a single active site at $t=0$.  ``Compact"
means that the density in surviving colonies remains finite
as $t \rightarrow \infty$.)

A closely related example is the {\em voter model}.\cite{durrett,liggett}
In this continuous-time Markov process sites of {\bf Z}$^d$ are either
in state 0 or state 1.  If site i is in state 0, it changes to 1 at rate
$\lambda r_1$, where $r_1$ is the number of nearest neighbors
in state 1.
(Note that only sites at cluster boundaries can change state.)
Similarly, sites in state 1 change to 0 at rate $r_0$.
(The case with $\lambda \neq 1$ is usually called the {\em biased}
voter model, for obvious reasons.)  Clearly, $\delta_{\phi}$ ($\delta_1$)
is the attractive
stationary state for $\lambda < 1$, ($\lambda > 1$), so
$\rho_1$ jumps from 0 to 1 at $\lambda = 1$.  The one-dimensional
voter model is essentially a continuous-time version of CDP with
$p_1 = \lambda/(1 + \lambda)$.
Thus we expect all results for critical exponents in the DKCA to
apply as well to the voter model in one dimension.

We can analyze spreading in the (unbiased) voter model in $d \geq 2$ as
follows.
Let $n_s$ be the number of 1's after $s$ changes in the configuration;
$n_0 = 1$.
$n_s$ is again a simple random walk on {\bf Z}, with an absorbing boundary at
the origin, so we know that the survival probability $\sim s^{-1/2}$, the
mean-square displacement over surviving walks
$\sim s$, and the mean displacement over all
walks is {\em O}(1).  The latter implies $\eta = 0$.
The number of steps per unit
time is proportional to the boundary
(number of 0-1 nearest neighbor pairs), $b_s$:

\begin{equation}
\frac{dt}{ds} \sim \frac{1}{b_s}.
\end{equation}

\noindent In one dimension, $b_s = 2$, so $t \sim s$.
For $d \geq 2$ the boundary
depends on the shape and internal structure of a colony;
we assume
$b_s \sim n^{\gamma}_s$ as $s \rightarrow \infty$.
Then since $n_s \sim s^{1/2}$, we have
$t(s) \sim s^{1-\gamma/2} $,  $n_{surv} (t) \sim t^{1/(2 -\gamma)}$,
and $P(t) \sim t^{-1/(2-\gamma)}$, so that $\delta = 1/(2-\gamma)$.
Since $b \leq qn$ on a lattice with coordination number $q$,
$\gamma \leq 1$, implying that $ \frac {1}{2} \leq \delta \leq 1$.

We expect asymptotic properties to be captured by a continuum description,
which for the voter model takes the form of a very
simple stochastic partial differential equation,
\begin{equation}
\frac {\partial \rho(x,t)}{\partial t} = \nabla^2 \rho (x,t) + \eta (x,t),
\label{field}
\end{equation}

\noindent where $\rho (x,t) $ is a coarse-grained
density of 1's and the (Gaussian) noise has autocorrelation
$<\eta(x,t)\eta(x',t')> = \delta(x-x') \delta (t-t') \rho (x,t) [1-\rho
(x,t)]$.
(There are no fluctuations when the density
is pinned at one of the absorbing values.)
In mean-field approximation (neglecting the noise)
the population spreads diffusively and $z=1$.
But we know that $z=1$ even for $d = 1$, suggesting the mean-field
exponent is correct for {\em all} $d$.

Simulations of the  voter model in two and three dimensions
support this conjecture.  In 2-$d$, a study of
$2 \times 10^{6}$
independent realizations (all starting with a single 1),
up to a maximum time of $10^4$, yield
$\delta = 0.95(3) $, $\eta = -0.01(1)$, and $z = 0.99(2)$
(figures in parentheses indicate
statistical uncertainties).  In three dimensions ($ 10^6$
realizations up to $t = 900$), we obtain $\delta \simeq 0.96(2)$,
$\eta = 0.00(1) $, and $z = 1.00(1)$.   (In both two and three dimensions
the colony mean-square radius of gyration  grows linearly with time.)
Moreover, the boundary exponent $\gamma$ appears to be unity in each case.
In 2-$d$, $\gamma$ shows considerable time dependence, increasing
monotonically from 0.82 for $t=1000$ to 0.93 for $t=20,000$, the longest
study attempted.  In three dimensions our study of colonies surviving to
$t=900$ yields $\gamma = 0.975(4)$.
The simplest conclusion is that $\gamma = \delta = 1$ for
$d \geq 2$.  Then the hyperscaling relation, Eq(\ref{dhyper}) is violated
for $d>2$, marking
$d=2$ as a kind of upper critical dimension for spreading
in  voter model.  (Consistent as well with the slow approach to
asymptotic scaling found in the 2-$d$ simulations.)
The density within a colony $ \sim n_{surv} (t)/R_t^d \sim
t^{1-d/2} $ when $\eta = 0 $ and $\delta = z = 1$, so
for $d > 2$ colonies are not compact, and
Eq(\ref{dhyper}) is not expected to hold.

In summary, we have shown how hyperscaling applies to critical spreading
at discontinuous transitions, and discussed applications to compact directed
percolation, and the closely-related voter model.

We are grateful to N. Inui, M. Katori, and H. Takayasu for helpful discussions.


\begin{thebibliography}{999}
\bibitem{bnote} Note that previous authors use $\beta$ to denote the survival
probability
exponent (our $\beta '$).  We prefer to reserve $\beta$ for
the order parameter exponent.
\end{thebibliography}
\end{document}